\documentclass[aps,twocolumn,showpacs,superscriptaddress]{revtex4-1}

\usepackage{graphicx,epsfig}
\usepackage{amsfonts,amsmath}
\usepackage{mathrsfs}
\usepackage{color}

\usepackage[utf8]{inputenc}

\def \sm {\sigma}
\def \dag {\dagger}
\def \ttg {\text{g}}
\def \veps {\varepsilon}

\begin{document}
\title{Coulomb-blockade effect in nonlinear mesoscopic capacitors}

\author{M. I. Alomar}
\affiliation{Institut de F\'{\i}sica Interdisciplin\`aria i Sistemes Complexos
IFISC (CSIC-UIB), E-07122 Palma de Mallorca, Spain}
\affiliation{Departament de F\'{\i}sica, Universitat de les Illes Balears, E-07122 Palma de Mallorca, Spain}

\author{Jong Soo Lim}
\affiliation{School of Physics, Korea Institute for Advanced Study, Seoul 130-722, Korea}

\author{David S\'anchez}
\affiliation{Institut de F\'{\i}sica Interdisciplin\`aria i Sistemes Complexos
IFISC (CSIC-UIB), E-07122 Palma de Mallorca, Spain}
\affiliation{Departament de F\'{\i}sica, Universitat de les Illes Balears, E-07122 Palma de Mallorca, Spain}

\begin{abstract}
We consider an interacting quantum dot working as a coherent source of single electrons.
The dot is tunnel coupled to a reservoir and capacitively coupled to a gate terminal
with an applied ac potential. At low frequencies, this is the quantum analog of 
the $RC$ circuit with a purely dynamical response. We investigate the quantized dynamics
as a consequence of ac pulses with large amplitude. Within a Keldysh-Green function formalism
we derive the time dependent current in the Coulomb blockade regime.
Our theory thus extends previous models that considered either noninteracting electrons in nonlinear response
or interacting electrons in the linear regime.
We prove that the electron emission and absorption resonances undergo a splitting
when the charging energy is larger than the tunnel broadening. For very large charging energies,
the additional peaks collapse and the original resonances are recovered, though with a reduced
amplitude.
Quantization of the charge emitted by the capacitor is reduced due to Coulomb repulsion
and additional plateaus arise.
Additionally, we discuss the differential capacitance and resistance as a function of time.
We find that to leading order in driving frequency the current can be expressed as a weighted
sum of noninteracting currents shifted by the charging energy.
\end{abstract}


\maketitle

\section{Introduction}\label{Sec:Intro}

Real-time manipulation of electrons is one of the greatest achievements
in modern nanoelectronics~\cite{gab06,fev07,gab12}.
The characteristic setup comprises a submicron-sized cavity or quantum dot tunnel coupled
to a reservoir through a quantum point contact. Then, a time dependent driving
voltage is applied to a electrostatically coupled
metallic gate placed on top of the dot. As a consequence,
dc transport is impossible and the system response is purely dynamical. The
low-frequency admittance measured with cryogenic low-noise amplifiers
can be understood from the serial combination
of a charge relaxation resistance and a quantum capacitance~\cite{but93,pre96}. It turns out that in the linear regime
(small ac amplitudes) the charge relaxation resistance is quantized for a single spin-polarized channel~\cite{but93a},
a theoretical prediction that was experimentally confirmed~\cite{gab06}. For drivings with larger amplitudes
(nonlinear regime), the system
works as an on-demand single-electron source~\cite{fev07}, in analogy with single-photon sources~\cite{mic00,san01},
with alternate sequences of electron emission
and absorption during a driving period in the fast (GHz) regime.
When the voltage pulse has a Lorentzian shape~\cite{lev96,iva97,kee06,dub13a},
recent progress has shown that the holes can be efficiently removed from the stream of excitations when the pulse is applied to an Ohmic contact~\cite{dub13}.
These phenomena imply the observation of quantized currents ensured by charge quantization, which might be useful
in metrology applications~\citep{pek13} and quantum computation designs~\cite{olk08,spl09,she12}.

Now, tunneling electrons feel repulsive interactions that yield Coulomb blockade, a prominent effect in small-capacitance
conductors which manifests itself as an increased resistance of a quantum dot junction at finite bias voltages~\cite{grabert}.
In fact, the effect is quite ubiquitous in nanoscale systems and arises not only in quantum dots but also in carbon
nanotubes~\cite{pos01}, molecular transistors~\cite{par02}, and optical lattices~\cite{che08}.
Therefore, it is natural to investigate the role of Coulomb blockade effects in single-electron sources.
This is the goal we want to accomplish in this work. We begin by noticing that electron-electron
interactions have been widely analyzed in the quantum $RC$ circuit~\cite{nig06,but07,nig08,rin08,rod09,mor10,ham10,spl10,alb10,fil11,lee11,fil12,con12,kas12,dut13,lim13,ros15,bur15}.
However, these works have mostly focused on the linear regime (for an exception, see Ref.~\cite{kas12}).
The nonlinear regime is interesting because both the capacitance and the charge relaxation resistance
acquire an explicit time dependence~\cite{mos08}. This result was found for noninteracting electrons.
Here, we give full expressions for the capacitive and the dissipative parts of the current
valid in the case of strong interactions that lead to Coulomb blockade effect. We predict that this effect
should be visible as a splitting of the dynamical current peaks for both emitted and absorbed electrons.
Importantly, the simultaneous emission of pairs of electrons in the non-interacting case is modified 
to a subsequent emission of two electrons.

\begin{figure}[t]
\centering
\includegraphics[width=0.40\textwidth]{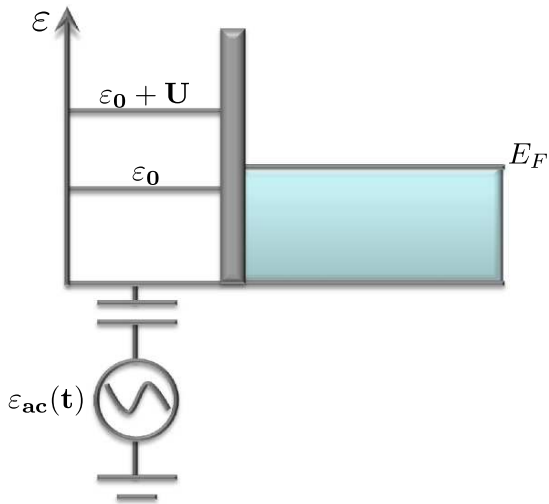}
\caption{Schematic representation of a single-electron source comprising a single-level quantum dot coupled capacitively to an ac oscillating signal, $\veps_{ac}(t)$. The dot can exchange electrons with an attached reservoir (Fermi energy $E_F$) via a tunnel barrier. 
The dot energy level is denoted with $\veps_0$ and Coulomb repulsion is given by the charging energy $U$.}
\label{Fig:Graph}
\end{figure}

The energy diagram of our system is sketched in Fig.~\ref{Fig:Graph}. We consider a single-level quantum dot
(energy $\veps_0$) coupled to a Fermi sea of electrons (Fermi energy $E_F$).
The coupling region between the dot and the reservoir
is typically a pinched-off quantum point contact that we depict in Fig.~\ref{Fig:Graph}
with a tunnel barrier. This part represents the resistive component of the quantum circuit, through which
electrons can hop on and off the dot. The position of $\veps_0$ can be tuned with a dc gate potential
applied to the point contact~\cite{fev07} (not shown in Fig.~\ref{Fig:Graph}). Additionally,
the dot is coupled to a nearby gate terminal with an externally applied harmonic potential $\veps_{ac}(t)$.
This is the capacitive part of the $RC$ circuit. Finally, a charging energy $U$ is required to charge
the dot with two electrons having opposite spins. The situation considered in this paper
is experimentally relevant for small dots.
The case of large dots with many quantum levels was treated in Ref.~\cite{nig06}, where a Hartree-Fock
approximation was employed to account for Coulomb interactions and screening effects.
Here, we consider the Anderson model with a single level and a constant interaction energy.
This model has been successfully applied to the Fermi liquid limit connected
to the Korringa-Shiba relation~\cite{fil11}, unveiling strong departures of the charge relaxation
resistance from universality~\cite{lee11}.

\section{Model Hamiltonian and Keldysh-Green function formalism}\label{Sec:Model}

Our theoretical discussion starts with the Anderson Hamiltonian of a mesoscopic capacitor, $H=H_R+H_T+H_{D}$, where $H_R$ describes the single reservoir, $H_T$ is the tunnel coupling between the reservoir and the quantum dot (QD) and $H_D$ models the QD: 
\begin{subequations}
\begin{eqnarray}\label{Eq:H}
H_R&=&\sum_{k\sm}\veps_{k} c_{k\sm}^{\dag} c_{k\sm}\,,\label{Eq:HR}\\
H_T&=&\sum_{k\sm}\left(V_{k}^{\ast} d_{\sm}^{\dag} c_{k\sm}+V_{k} c_{k\sm}^{\dag} d_{\sm}\right) \label{Eq:HT}\,,\\
H_D&=&\sum_{\sm}\veps_{\sm}(t) d_{\sm}^{\dag}d_{\sm}+U n_{\uparrow} n_{\downarrow} \label{Eq:HD}\,,
\end{eqnarray}
\end{subequations}
with $n_{\sm}=d_{\sm}^{\dag}d_{\sm}$ the occupation number operator and $\veps_{\sm}(t)=\veps_{\sm}+\veps_{ac}(t)$ 
including both the QD energy level, $\veps_{\sm} = \veps_0 + \sm\Delta_Z/2$ (here $\Delta_Z$ denotes the Zeeman splitting due to interaction with an external magnetic field), and the oscillating potential
applied to the gate, $\veps_{ac}(t)=\veps_{ac} \cos \Omega t$, where $\veps_{ac}$ is the
ac amplitude and $\Omega$ is the driving frequency. We emphasize that $\veps_{0}$ and $\veps_{ac}(t)$
can be tuned independently, as experimentally demonstrated~\cite{fev07}, with a dc and ac voltage, respectively,
applied to the quantum point contact and the gate electrode: $\veps_0=-eV_{\rm QPC}$ and $\veps_{ac}=-eV_g$.
This allows us to treat the position of the QD
level relative to the Fermi energy and the ac amplitude as separate parameters in our calculations. The sinusoidal drive considered here is convenient because the derivative of the drive is proportional to the frequency and thus easily Fourier decomposed. Different drives such as a step function do not shows this nice property and add mathematical difficulties to the formalism. Hence, we restrict ourselves to the monochromatic case.

In the Hamiltonian $H$, $\sm$ labels the electron spin and hereafter we consider the nonmagnetic case ($\Delta_Z = 0$).
However, the magnetic ($\Delta_Z \ne 0$) situation can be
easily included in our model but we focus on the spin-degenerate case. This is an important difference with the samples of Refs.~\cite{gab06,fev07},
which operate in the quantum Hall regime to achieve single-channel propagation with no spin degeneracy.

In Eq.~\eqref{Eq:HR}, $\veps_{k}$ represents the reservoir energy dispersion with momentum $k$
and $c_{k\sm}^{\dag}(c_{k\sm})$ creates (annihilates) a conduction band electron.
The tunnel hamiltonian given by Eq.~\eqref{Eq:HT} contains the tunnel amplitude $V_{k}$ and the fermionic operator
$d_{\sm}^{\dag}(d_{\sm})$, which creates (annihilates) a localized electron in the dot.
Finally, $U=e^2/(C_g+C_R)$ in Eq.~\eqref{Eq:HD} is the charging energy, which we also take as a tunable parameter depending
on the capacitive strengths with the coupled gate, $C_g$, and eventually with the reservoir, $C_R$.

The time dependent field $\veps_{ac}(t)$ induces a purely dynamical charge current $I_R(t)$ that can be measured
at the reservoir. Since $H$ commutes with the total charge, $I_R$ is determined from the change rate of the dot
occupation, $I(t)$:
\begin{equation}\label{Eq:Cont}
I_R(t)+I(t)=0\,,
\end{equation}
where $I(t)=e \partial_t\sum_{\sm} \langle d_{\sm}^{\dag} d_{\sm}\rangle (t)$
and $I_R(t)=e \partial_t\sum_{k,\sm} \langle c_{k\sm}^{\dag} c_{k\sm}\rangle (t)$ with $e$ the unit of charge.
Here, $\partial_t$ denotes the time derivative. Equation~\eqref{Eq:Cont} thus represents the electronic charge conservation. 
In what follows, we focus on $I(t)$ because it can be directly expressed
in terms of the QD Green's function without further manipulation, as shown below. 
The physical current $I_R$ (since it amounts to a flux) can then be obtained immediately from Eq.~\eqref{Eq:Cont}.

Let $G_{\sm}^< (t,t')=i\langle d_{\sm}^{\dag}(t') d_{\sm}(t)\rangle$ be the lesser Green's function~\cite{jau94,jauho}
for the dot operators.
Clearly, the QD occupation $\langle n_{\sm}(t)\rangle=\langle d_{\sm}^{\dag}(t)d_{\sm}(t)\rangle $
can be written in terms of the lesser Green's function. The current is hence calculated as
\begin{align}\label{Eq:I(t)}
I(t) &=e \partial_t\sum_{\sm} \left\langle n_{\sm}(t)\right\rangle 
= e\partial_t\sum_{\sm}\Big( -i {G}_{\sm}^< (t,t)\Big) 
\nonumber \\
&= e\partial_t\sum_{\sm}\int \frac{d\veps}{2 \pi i}{G}_{\sm}^<(t,\veps)\,,
\end{align}
where in the last line we express the lesser dot Green's function in a mixed time energy notation~\cite{Ara05,Ara06}.
This representation is especially useful for nonstationary scattering problems in the adiabatic limit~\cite{mos09,moskalets}.
Its connection with the original double time picture and the corresponding Fourier transform is discussed in
Appendix~\ref{App:Fourier}.

Our regime of interest here is the adiabatic case (small frequency $\Omega$)
but arbitrary values of the ac amplitude $\veps_{ac}$. In that case, the Green's function
is expected to display small deviations around a frozen state in time characterized by a stationary
scattering matrix with time dependent parameters. This approximation is good when $\hbar\Omega$
is the smallest energy scale of our problem. For a prototypical $RC$ circuit~\cite{gab06},
$\hbar\Omega\simeq 0.2$~$\mu$eV, which is
at least fifty times smaller than the tunnel coupling $\Gamma\simeq 10$~$\mu$eV.
Therefore, the electron interacts only weakly with the ac potential before tunneling into or out of the QD.
The frequency expansion reads,
\begin{align}\label{Eq:Frec}
{G}_{\sm}^<(t,\veps)={G}_{\sm}^{<,f}(t,\veps)+\hbar\Omega{G}_{\sm}^{<,(1)}(t,\veps) + {O}(\Omega^2)\,,
\end{align}
where the superscript $f$ denotes the frozen approximation and $(1)$ implies the first order in driving frequency $\Omega$. 
Second-order terms and beyond are neglected, which
suffices for the purposes of this work. (Inductive-like effects have been studied in Ref.~\cite{guo}).
We stress that the zeroth-order (frozen) term in $\Omega$
is still time dependent. No assumption has been made on the strength of the amplitude,
which can be arbitrarily large, driving the system into the nonlinear regime.

Substituting Eq.~\eqref{Eq:Frec} into Eq.~\eqref{Eq:I(t)},
we find similar expansions for the occupation and the current,
\begin{align}\label{Eq:I(t)12}
I(t)&\simeq e\partial_t\sum_{\sm}\int \frac{d\veps}{2 \pi i}\Big({G}_{\sm}^{<,f}(t,\veps)+\hbar\Omega{G}_{\sm}^{<,(1)}(t,\veps)\Big) 
\nonumber\\
&=e \partial_t\sum_{\sm}(\left\langle n_{\sm}(t)\right\rangle ^{f}+\left\langle n_{\sm}(t)\right\rangle ^{(1)}) 
\nonumber\\
&=I^{(1)}(t)+I^{(2)}(t)\,.
\end{align}
From the definition given by Eq.~\eqref{Eq:I(t)}, it follows that the leading order
for the current is first order in $\Omega$. To be consistent, we therefore keep the current terms
in Eq.~\eqref{Eq:I(t)12} up to second order in $\Omega$.
The physical implication says that $I^{(1)}$ represents a capacitive-like contribution while
$I^{(2)}$ is understood as a dissipative component~\cite{mos08}.

This interpretation can be substantiated by introducing a quantum $RC$ circuit model
(a capacitor and a resistor in a series with an applied ac potential)
with time dependent capacitance and resistance functions,
\begin{multline}\label{Eq:eI(t)}
eI(t)\simeq-C_{\partial}(t)\partial_t \veps_{ac}(t)\\
+R_{\partial}(t)C_{\partial}(t)\partial_t(C_{\partial}(t)\partial_t \veps_{ac}(t)) \,.
\end{multline}
This relation is valid at low frequency for both the linear and the nonlinear regimes.
In Eq.~\eqref{Eq:eI(t)} $C_{\partial}(t)$ is the differential capacitance and
$R_{\partial}(t)$ the differential resistance. Both depend on time because
they constitute a generalization of the linear-response quantum capacitance $C_q$ 
and charge relaxation resistance $R_q$~\cite{but93} to the nonlinear ac transport regime~\cite{mos08}.
Combining Eq.~\eqref{Eq:I(t)12} with Eq.~\eqref{Eq:eI(t)}, we can find expressions
for $C_{\partial}(t)$ and $R_{\partial}(t)$. Therefore, our goal is first to obtain
an equation for $G^<_{\sm}(t,\veps)$ in the presence of Coulomb interactions
and oscillating voltages.

\section{Equation of motion}\label{Sec:Equation}

The temporal evolution of the dot Green's function is determined from the commutator
of $d_\sm$ with $H$ (Heisenberg equation of motion).
It is convenient to consider the time-ordered Green's function $G_{\sm}(t,t') \equiv \langle \langle d_{\sm},d_{\sm}^{\dag}\rangle\rangle (t,t') = -i\langle{\cal{T}}d_{\sm}(t)d_{\sm}^{\dag}(t')\rangle$.
After some straightforward steps, we find that the time-ordered Green's function satisfies the integral (Dyson) equation
\begin{multline}\label{Eq:G}
G_{\sm}(t,t')=\ttg_{\sm}(t,t')+\int\frac{ds}{\hbar}~G_{\sm}(t,s)\veps_{ac}(s) \ttg_{\sm}(s,t')
\\
+\int\frac{ds}{\hbar}~\int\frac{ds'}{\hbar}~G_{\sm}(t,s') \Sigma_0(s',s) \ttg_{\sm}(s,t')
\\
+U\int\frac{ds}{\hbar}~\langle\langle d_{\sm},d_{\sm}^{\dag}n_{\bar{\sm}}\rangle\rangle(t,s) \ttg_{\sm}(s,t') \,,
\end{multline}
where $\bar{\sm}=-\sm$.
$\Sigma_0(t,t')= \sum_{k}\vert V_{k}\vert^2 \ttg_{k}(t,t')$ is the tunnel self-energy with $\ttg_{k(\sm)}(t,t')$ the isolated reservoir (dot) Green's function in the absence of the ac driving potential. 
The retarded/advanced and lesser Green's functions can then be obtained from the Langreth's analytic continuation rules~\cite{jauho}.

To consider the effect of $U$, we now generate an additional integral equation for the correlator $\langle\langle   d_{\sm},d_{\sm}^{\dag}n_{\bar{\sigma}}\rangle\rangle(t,t^{\prime})$ in Eq.~\eqref{Eq:G}:
\begin{multline}\label{Eq:Corr.1}
\langle\langle d_{\sm},d_{\sm}^{\dag}n_{\bar{\sm}}\rangle\rangle(t,t')
=
\left\langle n_{\bar{\sm}}(t)\right\rangle \ttg_{\sm}(t,t')
\\
+\int\frac{ds}{\hbar}~ \langle\langle d_{\sm},d_{\sm}^{\dag}n_{\bar{\sm}}\rangle\rangle(t,s)\veps_{ac}(s) \ttg_{\sm}(s,t')
\\
+ \sum_k\int\frac{ds}{\hbar}~\left(V_{k}\langle\langle d_{\sm},c^{\dag}_{k\sm}n_{\bar{\sm}}\rangle\rangle(t,s)\right. 
\\
\left. +V_{k}\langle\langle d_{\sm},d_{\sm}^{\dag}c_{k\bar{\sm}}d_{\bar{\sm}}\rangle\rangle(t,s)\right. \\
\left. -V_{k}^{\ast}\langle\langle d_{\sm},d_{\sm}^{\dag}d_{\bar{\sm}}^{\dag}c_{k\bar{\sm}}\rangle\rangle(t,s)\right)\ttg_{\sm}(s,t')
\\
+\,U\int\frac{ds}{\hbar}~\langle\langle d_{\sm},d_{\sm}^{\dag}n_{\bar{\sm}}\rangle\rangle(t,s) \ttg_{\sm}(s,t')\,,
\end{multline}
where three new correlation functions arise. Since we are interested in the Coulomb blockade regime, we can neglect charge and spin excitations. This truncated equation of motion approach is good in the weak tunneling regime or for not very low temperatures, in which
case Kondo correlations can be disregarded~\cite{Van10}. As a consequence, we neglect the spin-flip correlators in Eq.~\eqref{Eq:Corr.1}:
\begin{subequations}\label{Eq:Aprox.1}
\begin{align}
&\langle\langle   d_{\sm},d_{\sm}^{\dag}d_{\bar{\sm}}^{\dag}c_{k\bar{\sm}}\rangle\rangle(t,s)\simeq 0\,,\\
&\langle\langle   d_{\sm},d_{\sm}^{\dag}c_{k\bar{\sm}}d_{\bar{\sm}}\rangle\rangle(t,s)\simeq 0\,.
\end{align}
\end{subequations}

Next, we calculate the equation of motion for $\langle\langle d_{\sm},c^{\dag}_{k\sm}n_{\bar{\sm}}\rangle\rangle(t,t')$:
\begin{multline}\label{Eq:Corr.2}
\langle\langle d_{\sm},c^{\dag}_{k\sm}n_{\bar{\sm}}\rangle\rangle(t,t')=
\int\frac{ds}{\hbar}~ V^{\ast}_{k}\langle\langle d_{\sm},d_{\sm}^{\dag}n_{\bar{\sm}}\rangle\rangle(t,s) \ttg_{k}(s,t')\\
+\sum_k\int\frac{ds}{\hbar}~\left(V_{k}\langle\langle d_{\sm},c^{\dag}_{k\sm}c^{\dag}_{k\bar{\sm}}d_{\bar{\sm}}\rangle\rangle(t,s)\right. 
\\
\left. -V_{k}^{\ast}\langle\langle d_{\sm},c^{\dag}_{k\sm}d^{\dag}_{\bar{\sm}}c_{k\bar{\sm}}\rangle\rangle(t,s)\right)\ttg_{k}(s,t') \,,
\end{multline}
where we neglect reservoir charge and spin excitations for the same reason as discussed above,
\begin{subequations}\label{Eq:Aprox.2}
\begin{align}
&\langle\langle d_{\sm},c^{\dag}_{k\sm}c^{\dag}_{k\bar{\sm}}d_{\bar{\sm}}\rangle\rangle
(t,s)\simeq 0\,,\\
&\langle\langle d_{\sm},c^{\dag}_{k\sm}d^{\dag}_{\bar{\sm}}c_{k\bar{\sm}}\rangle\rangle(t,s)\simeq 0\,.
\end{align}
\end{subequations}

Combining Eqs.~\eqref{Eq:Corr.1} and~\eqref{Eq:Corr.2} with Eqs.~\eqref{Eq:Aprox.1} and~\eqref{Eq:Aprox.2} we obtain a closed expression for $\langle\langle   d_{\sm},d_{\sm}^{\dag}n_{\bar{\sigma}}\rangle\rangle(t,t^{\prime})$:
\begin{multline}\label{Eq:Corr}
\langle\langle   d_{\sm},d_{\sm}^{\dag}n_{\bar{\sm}}\rangle\rangle(t,t')=\left\langle n_{\bar{\sm}}(t)\right\rangle \ttg_{\sm}(t,t')
\\
+\int\frac{ds}{\hbar}~ \langle\langle d_{\sm},d_{\sm}^{\dag}n_{\bar{\sm}}\rangle\rangle(t,s)\veps_{ac}(s) \ttg_{\sm}(s,t')
\\
+\int\frac{ds}{\hbar}~\int\frac{ds'}{\hbar}~ \langle\langle d_{\sm},d_{\sm}^{\dag}n_{\bar{\sm}}\rangle\rangle(t,s') \Sigma_0(s',s) \ttg_{\sm}(s,t')
\\
+\,U\int\frac{ds}{\hbar}~\langle\langle d_{\sm},d_{\sm}^{\dag}n_{\bar{\sm}}\rangle\rangle(t,s) \ttg_{\sm}(s,t') \,.
\end{multline}
We have thus derived two coupled integral equations, namely Eqs.~\eqref{Eq:G} and~\eqref{Eq:Corr},
which must be self-consistently solved because Eq.~\eqref{Eq:Corr} depends on $\left\langle n_{{\sm}}(t)\right\rangle$ and to calculate this quantity we need to know $G^<_{{\sigma}}(t,t)$ (see Eq.~\eqref{Eq:G}), which depends itself on $\left\langle n_{{\sm}}(t)\right\rangle$ via Eq.~\eqref{Eq:Corr}. 
Further progress can be made by expanding the equations in powers of driving frequency $\Omega$. 
It is worthwhile to emphasize that Eq.~\eqref{Eq:G} is exact while Eq.~\eqref{Eq:Corr} is a quite reasonable approximation that works fairly well in the Coulomb blockade regime.

\section{Noninteracting case}\label{Sec:NonInt.}

It is instructive to begin our discussion with the independent particle approximation.
This is easy to accomplish by setting $U=0$
in Eq.~\eqref{Eq:G}. Thus, we obtain an integral equation that depends on the dot Green's function only,
\begin{multline}\label{Eq:G0}
\mathcal{G}_{\sm}(t,t')=\ttg_{\sm}(t,t')+\int\frac{ds}{\hbar}~\mathcal{G}_{\sm}(t,s)\veps_{ac}(s) \ttg_{\sm}(s,t')
\\
+\int\frac{ds}{\hbar}~\int\frac{ds'}{\hbar}~\mathcal{G}_{\sm}(t,s') \Sigma_0(s',s) \ttg_{\sm}(s,t') \,.
\end{multline}
Importantly, we have changed our notation $G\rightarrow\mathcal{G}$ in order to distinguish between the Green's function corresponding to the the Coulomb Blockade regime ($G$) and that for noninteracting electrons ($\mathcal{G}$).
This is done for later convenience since we will show that interacting results can indeed be expressed
using noninteracting quantities.

A frequency expansion of Eq.~\eqref{Eq:G0} yields (we refer the reader to Appendix~\ref{App:Serie.0} for details):
\begin{subequations}\label{Eq:Gs0}
\begin{align}
&\mathcal{G}_{\sm}^{r/a,f}(t,\veps)=\frac{1}{\veps-\veps_{\sm}-\veps_{ac}(t)-\Sigma_0^{r/a}(\veps)}\,,\label{Eq:Grf0}
\\
&\mathcal{G}_{\sm}^{r/a,(1)}(t,\veps)=\frac{i}{\Omega}\partial_t\veps_{ac}(t)\mathcal{G}_{\sm}^{r/a,f}(t,\veps)\partial_{\veps} \mathcal{G}_{\sm}^{r/a,f}(t,\veps) \,,\label{Eq:Gr(1)0}
\\
&\mathcal{G}_{\sm}^{<,f}(t,\veps)=\mathcal{G}_{\sm}^{r,f}(t,\veps)\Sigma^{<}_0(\veps)\mathcal{G}_{\sm}^{a,f}(t,\veps)\,,\label{Eq:G<f0}
\\
&\mathcal{G}_{\sm}^{<,(1)}(t,\veps)=\frac{i}{\Omega}\partial_t\veps_{ac}(t)\left( \mathcal{G}_{\sm}^{a,f}(t,\veps)\partial_{\veps}\mathcal{G}_{\sm}^{<,f}(t,\veps)\right.\nonumber 
\\
&\left.\quad\quad\quad\quad\quad\quad\quad\quad\quad\quad +\,\mathcal{G}_{\sm}^{<,f}(t,\veps)\partial_{\veps}\mathcal{G}_{\sm}^{r,f}(t,\veps)\right) \,,\label{Eq:G<(1)0}
\end{align}
\end{subequations}
where the superscript ``$r/a$'' labels the retarded/advanced Green's function and the tunnel self-energies read
$\Sigma_0^{r/a}(\veps)=\mp i\Gamma$,  $\Sigma_0^<(\veps)=2i\Gamma f(\veps)$.
$\Gamma=\pi |V_k|^2\rho$ is the hybridization width, which we take as a constant parameter.
This is a good approximation when the tunnel probability $|V_k|^2$ and the lead density of states
$\rho$ depend weakly on energy, which is the experimentally relevant situation.
$f(\veps)=1/[1+\exp{(\veps-E_F)/k_B T}]$ denotes the Fermi-Dirac distribution
with $E_F$ the lead Fermi level and $T$ the base temperature.

We consider the spin-degenerate case ($\Delta_Z = 0$). Therefore, the dot level fulfills
\begin{align}\label{Eq:nonmagnetic}
\veps_{\uparrow}=\veps_{\downarrow}\equiv\varepsilon_0\,,
\end{align}
and we can define a total dot occupation $\left\langle n(t)\right\rangle^f_0$ as
\begin{align}
\left\langle n_{\uparrow}(t)\right\rangle^f_0=\left\langle n_{\downarrow}(t)\right\rangle^f_0 \equiv \left\langle n(t)\right\rangle^f_0/2\,.
\end{align}
Here, the subscript $0$ means ``noninteracting''.
Using the expressions for the noninteracting Green's functions given by
Eqs.~(\ref{Eq:Gs0}), the current and mean occupation
implied by Eq.~\eqref{Eq:I(t)12} become
\begin{subequations}\label{Eq:NI0}
\begin{align}
&\left\langle n(t)\right\rangle^f_{0}= 2\int d\veps~ f(\veps)\mathcal{D}(t, \veps) \,,\label{Eq:nf0}
\\
&I^{(1)}_0(t)=-2e\int d\veps~ \left(-\partial_{\veps} f(\veps)\right) \mathcal{D}(t,\veps) \partial_t \veps_{ac}(t) \,,\label{Eq:I10}
\\
&\left\langle n(t)\right\rangle^{(1)}_{0} = h\int d\veps~ \left(-\partial_{\veps} f(\veps)\right)\mathcal{D}^2(t, \veps)\partial_t \veps_{ac}(t) \,,\label{Eq:n10}
\\
&I^{(2)}_0(t)=eh\int d\veps~ \left(-\partial_{\veps} f(\veps)\right)\partial_t\big(\mathcal{D}^2(t, \veps)\partial_t \veps_{ac}(t)\big) \,,\label{Eq:I20}
\end{align}
\end{subequations}
where $\mathcal{D}(t,\veps)\equiv\mathcal{D}_{\uparrow}(t,\veps)=\mathcal{D}_{\downarrow}(t,\veps)$
is the density of states written as
\begin{eqnarray}\label{Eq:Dte}
\mathcal{D}_{\sm}(t,\veps) =\frac{1}{\pi}\frac{ \Gamma}{\big(\veps-\veps_{\sm}-\veps_{ac}(t)\big)^2+\Gamma^2} \,.
\end{eqnarray}

Equation~\eqref{Eq:Dte} is a Breit-Wigner-like density of states which instantaneously changes with time.
This is a physically transparent result---in the adiabatic regime the dot spectral function is given
by the stationary density of states replacing the dot level $\veps_0$ with the instantaneous
variation of the dot potential as a function of time, i.e., $\veps_0\to\veps_0+\varepsilon_{ac}(t)$.
In other words, the electron adjusts its dynamics to the slow ac potential.
As a consequence, the frozen occupation [Eq.~\eqref{Eq:nf0}] is simply given by the integral of the local
density states convoluted with the Fermi function. The next order in the $\Omega$ expansion
[Eq.~\eqref{Eq:n10}] depends on the derivative of $\veps_{ac}(t)$, as it should.
For small frequencies, this is a small correction to the frozen occupation.
Finally, the capacitive and dissipative currents [Eqs.~\eqref{Eq:I10} and~\eqref{Eq:I20}]
are just given by time derivatives of the frozen and the first-order occupations, respectively.
At very low temperatures, the main contribution to both current contributions arises from the electrons
around the Fermi energy due to the $-\partial_{\veps} f$ term in the equations.

\begin{figure}[t]
\centering
\includegraphics[width=0.40\textwidth]{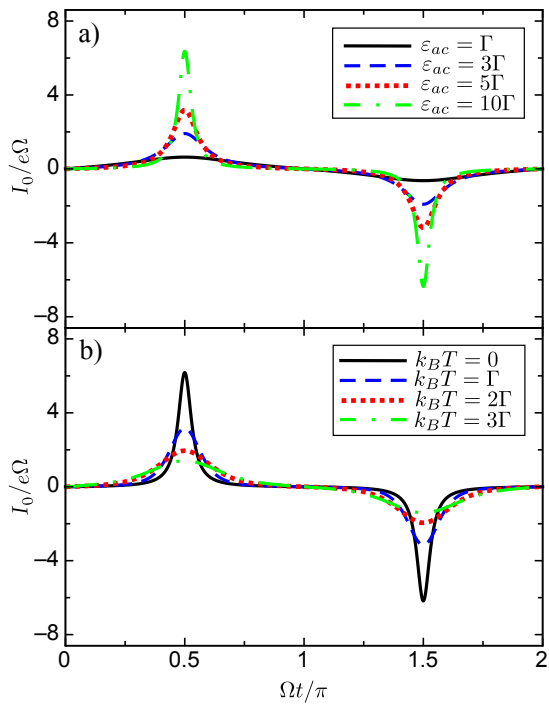}
\caption{Noninteracting charge current (up to second order in the ac frequency) as a function time for different ac amplitudes (a) and temperatures (b). Parameters: $\veps_0=0$, $\hbar\Omega=0.02\,\Gamma$, (a) $k_BT=0$, and (b) $\veps_{ac}=10\Gamma$.}
\label{Fig:I1kBTEac10EacEd0}
\end{figure}

\begin{figure}[b]
\centering
\includegraphics[width=0.46\textwidth]{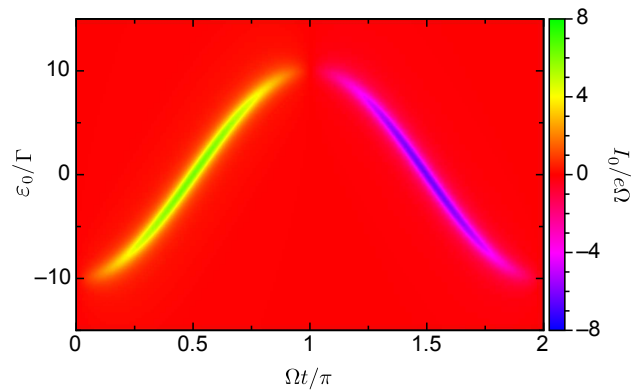}
\caption{Noninteracting charge current (up to second order in the ac frequency) 
as a function of dot energy level (vertical axis) and time (horizontal).
Parameters: $\veps_{ac}=10\Gamma$, $\hbar\Omega=0.02\,\Gamma$, and $k_BT=0$.}
\label{Fig:I1OtEdEac10}
\end{figure}

The total current $I_0(t)=I^{(1)}_0(t)+I^{(2)}_0(t)$ is plotted in Fig.~\ref{Fig:I1kBTEac10EacEd0}(a) as a function of time for different $\veps_{ac}$ amplitudes. The results are calculated for zero temperature and very small ac frequencies.
In the large amplitude case (green dashed-dotted line), we observe a current peak (dip) in the first (second) half cycle since in the first (second) half cycle an electron is adsorbed (emitted) by the dot. 
This occurs when the ac modulated dot level aligns with the Fermi level, $\veps_0+\veps_{ac}\cos \Omega t=E_F$ (hereafter we set $E_F = 0$). The amplitude of the current peak (dip) is proportional to $\varepsilon_{ac}$, as shown in Eqs.~\eqref{Eq:I10} and~\eqref{Eq:I20}.
Therefore, the ac amplitude should be larger than $\Gamma$ for the single-electron source to produce
well defined current peaks. This is within experimental reach since $\veps_{ac}\simeq 100$~$\mu$eV~\cite{fev07}
and $\Gamma\simeq 10$~$\mu$eV. On the other hand, the ac frequency should be
$\hbar\Omega=0.02\,\Gamma\simeq 0.2$~$\mu$eV
and the resulting current peak, given in Fig.~\ref{Fig:I1kBTEac10EacEd0}(a) in units of $e\Omega$,
attains values of the order of $I_0\simeq 0.3$~nA, which is experimentally measurable.

At nonzero temperatures, the peaks broaden due to thermal smearing
[see Fig.~\ref{Fig:I1kBTEac10EacEd0}(b)]. The reason is clear---for large temperatures
(larger than $\Gamma$) and fixed ac amplitude the current pulse is distributed among electronic
states within $k_BT$ around the Fermi energy and the pulse is not sharply peaked as in the $k_BT=0$
case. As a consequence, low temperatures smaller than $T\simeq 100$~mK (=8.62$\mu$eV) 
for $\Gamma\simeq 10$~$\mu$eV are needed to observe single-electron
injection into the Fermi sea.

Figure~\ref{Fig:I1OtEdEac10} shows the total current for a fixed $\veps_{ac}$ as a function of time
(horizontal axis) and the dot level position (vertical axis). The peak and dip found in  Fig.~\ref{Fig:I1kBTEac10EacEd0} are also visible in Fig.~\ref{Fig:I1OtEdEac10} within a value range of $\veps_0$. 
The current resonances shift with time in order to satisfy the resonant condition $\veps_0+\veps_{ac}\cos \Omega t=E_F$. 
Notably, for dot levels such that $|\veps_0|>|\veps_{ac}|$ the current is identically zero independently of time, since at those energies the resonant condition is never met.

\begin{figure}[t]
\centering
\includegraphics[width=0.40\textwidth]{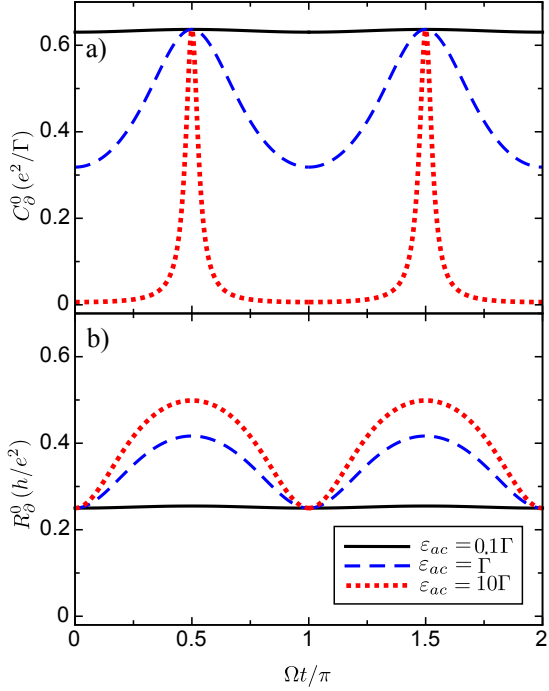}
\caption{Differential capacitance (a) and differential resistance (b) as a function of time for different ac amplitudes. Parameters: $\veps_0=0$, $\hbar\Omega = 0.02\Gamma$, and $k_BT=0$.}
\label{Fig:CqRqEac}
\end{figure}

Now, using Eqs.~\eqref{Eq:eI(t)}, \eqref{Eq:I10}, and~\eqref{Eq:I20} we derive the following expressions for the differential capacitance and resistance:
\small\begin{align}
& C^0_{\partial}(t)= 2e^2\int\!\! d\veps (-\partial_{\veps} f) \mathcal{D}(t,\veps) \,,\label{Eq:C0t}
\\
& R^0_{\partial}(t)=
\frac{h}{4 e^2} \frac{\int\!\! d\veps (-\partial_{\veps}f)\partial_t\big(\mathcal{D}^2(t,\veps)\partial_t \veps_{ac}(t)\big)}
{\int\!\! d\veps (-\partial_{\veps} f) \mathcal{D}(t,\veps) \!\!\int\!\! d\veps (-\partial_{\veps} f) \partial_t\big(\mathcal{D}(t,\veps)\partial_t\veps_{ac}(t)\big)} \,,
\label{Eq:R0t}
\end{align}
where the dot density of states $\mathcal{D}(t,\veps)$ is given by Eq.~\eqref{Eq:Dte}.
Clearly, Eq.~\eqref{Eq:C0t} can be interpreted as an instantaneous quantum capacitance.
The physical meaning of the resistance of Eq.~\eqref{Eq:R0t} is less obvious.
Only in linear response does $R^0_{\partial}(t)$ reduce to the charge relaxation resistance~\cite{mos08}.

Figure~\ref{Fig:CqRqEac}(a) shows $C^0_{\partial}(t)$ for three specific cases: $\veps_{ac}=0.1\Gamma$
(solid black line), $\Gamma$ (dashed blue line) and $10\Gamma$ (dotted red line). In the first case, $C^0_{\partial}$ is nearly time independent and takes on its maximum value as a constant times $e^2/\Gamma$. This occurs because
in the low $\veps_{ac}$ limit the dot density of states has a constant value for any time. As the ac amplitude increases, a strong time dependence becomes apparent in terms of two well defined peaks when the aforementioned resonant condition is fulfilled. We observe that the minima of the dashed blue line never reaches zero since for intermediate values of $\veps_{ac}$ the dot energy level is close to $E_F$ and can therefore be populated. In the strongly nonlinear case
(dotted red line) the two peaks become clearly resolved inasmuch as for large $\veps_{ac}$ the dot level gets fully depopulated (populated) after electron emission (injection).

In the linear regime ($\veps_{ac}\rightarrow 0$) and zero temperature the quantum capacitance given by Eq.~\eqref{Eq:C0t} takes a simpler form, $C^0_{\partial}=2 e^2 \mathcal{D}$, 
which is time independent and provides information about the dot density of states as we tune $\veps_0$. 
In fact, the static density of states becomes $\mathcal{D}=(\Gamma/\pi)/[(E_F-\veps_0)^2+\Gamma^2]$, i.e., a Lorentzian curve centered at $E_F$ with half-width $\Gamma$. Hence, the value marked by the solid black line of Fig.~\ref{Fig:CqRqEac}(a)
is not universal and depends on the position of $\veps_0$ with respect to $E_F$~\cite{but07}. In particular,
for $\veps_0=0$ the capacitance is $C^0_{\partial}=2 e^2/ \pi \Gamma\simeq 0.64 e^2/\Gamma $ as shown
in Fig.~\ref{Fig:CqRqEac}(a).
In contrast, the resistance in the linear regime and for $k_BT=0$ is not sample specific. $R^0_{\partial}$ becomes time and energy independent [see the solid black line of  Fig.~\ref{Fig:CqRqEac}(b)], 
taking the universal value $R^0_{\partial}={h}/{4 e^2}=0.25h/e^2$ (we recall that we have two independent channels, one per spin). 
This quantization of the resistance was earlier predicted by B\"uttiker \textit{et al.} in 1993~\cite{but93} and later demonstrated experimentally for the spin-polarized case by Gabelli \textit{et al.} in 2006~\citep{gab06}.
This resistance can be also connected with an instantaneous Joule law for the dissipated heat
in the reservoir~\cite{lud14,lud16}.

Away from linear response [dashed blue line and dotted red line in Fig.~\ref{Fig:CqRqEac}(b)],
the resistance quickly deviates from the quantized value and becomes both time and energy dependent.
With increasing $\veps_{ac}$, $R^0_{\partial}$ shows two peaks as a result of the resonant condition
but, unlike the capacitance, the resistance peaks get higher and more broadened as the ac amplitude increases.
Therefore, the dissipation enhances as $\veps_{ac}$ grows, which is naturally expected.
The enhancement rate is, however, nonlinear and not easily derived from Eq.~\eqref{Eq:R0t}.

\section{Coulomb Blockade Regime}\label{Sec:CB}

Our aim now is to include Coulomb repulsion between electrons in the quantum dot and to investigate how
the noninteracting results discussed in the previous section change in the presence of interactions.
In the Coulomb blockade regime, the charging energy is typically a large energy scale in the problem and for small
dots one has $U>\pi \Gamma$~\cite{grabert}. We start from the main results of the equation-of-motion
method [Eqs.~\eqref{Eq:G} and~\eqref{Eq:Corr}]. The frequency expansion can be performed after somewhat lengthy calculations detailed in Appendix~\ref{App:Serie.CB}. We find the frozen and dynamic (to leading order in $\Omega$) lesser and
retarded Green's functions,
\begin{widetext}
\begin{subequations}\label{Eq:GCB}
\begin{align}
&G^{r,f}_{\sm}(t,\veps)=\big(1-\left\langle n_{\bar{\sm}}(t)\right\rangle^f\big)\mathcal{G}^{r,f}_{\sm}(t,\veps)+\left\langle n_{\bar{\sm}}(t)\right\rangle^f\mathcal{G}^{r,f}_{U\sm}(t,\veps)\,,\label{Eq:Grfte}
\\
&G^{r,(1)}_{\sm}(t,\veps)= \Big( U\,\left\langle n_{\bar{\sm}}(t)\right\rangle^{(1)}\mathcal{G}^{r,f}_{U\sm}(t,\veps)+\frac{i}{\Omega}\partial_t\veps_{ac}(t)\Big[ \big(1-\left\langle n_{\bar{\sm}}(t)\right\rangle^f\big)\partial_{\veps} \mathcal{G}^{r,f}_{\sm}(t,\veps)\nonumber\\
&\quad\quad\quad\quad\quad\quad\quad\quad\quad\quad\quad\quad\quad\quad\quad\quad\quad\quad\quad\quad\quad\quad\quad\quad\quad\quad\quad\quad  +(1+U\,\mathcal{G}^{r,f}_{U\sm}(t,\veps))\left\langle n_{\bar{\sm}}(t)\right\rangle^f\partial_{\veps} \mathcal{G}^{r,f}_{U\sm}(t,\veps)\Big]\Big)\mathcal{G}^{r,f}_{\sm}(t,\veps)\,,\label{Eq:Gr(1)te}
\\
&G^{<,f}_{\sm}(t,\veps)=\big(1-\left\langle n_{\bar{\sm}}(t)\right\rangle^f\big)\mathcal{G}^{<,f}_{\sm}(t,\veps) +\left\langle n_{\bar{\sm}}(t)\right\rangle^f\mathcal{G}^{<,f}_{U\sm}(t,\veps)\,,\label{Eq:G<fte}
\\
&G^{<,(1)}_{\sm}(t,\veps)= \left\langle n_{\bar{\sm}}(t)\right\rangle^{(1)}\Big( \mathcal{G}^{<,f}_{\sm}(t,\veps)-\mathcal{G}^{<,f}_{U\sm}(t,\veps)\Big)+\frac{i}{\Omega}\partial_t\veps_{ac}(t)\Big( \big(1-\left\langle n_{\bar{\sm}}(t)\right\rangle^f\big)\Big[ \mathcal{G}^{a,f}_{\sm}(t,\veps)\partial_{\veps}\mathcal{G}^{<,f}_{\sm}(t,\veps) +\mathcal{G}^{<,f}_{\sm}(t,\veps)\partial_{\veps}\mathcal{G}^{r,f}_{\sm}(t,\veps)\Big]\nonumber\\
&\quad\quad\quad\quad\quad\quad\quad\quad\quad\quad\quad\quad\quad\quad\quad\quad\quad\quad\quad\quad\quad\quad\quad\quad\quad +\left\langle n_{\bar{\sm}}(t)\right\rangle^f\Big[ \mathcal{G}^{a,f}_{U\sm}(t,\veps)\partial_{\veps}\mathcal{G}^{<,f}_{U\sm}(t,\veps) +\mathcal{G}^{<,f}_{U\sm}(t,\veps)\partial_{\veps}\mathcal{G}^{r,f}_{U\sm}(t,\veps)\Big]\Big)\,.\label{Eq:G<(1)te}
\end{align}
\end{subequations}
\end{widetext}
Here, we express the interacting Green's functions (denoted by $G$) in terms of the noninteracting Green's functions
[denoted by $\mathcal{G}$ and explicitly written in Eqs.~(\ref{Eq:Gs0})].
We indicate with the subscript $U$ that $\mathcal{G}_{U\sm}$ is the noninteracting Green's function with the replacement $\veps_0\rightarrow \veps_0+U$.

We focus on the nonmagnetic case as in Sec.~\ref{Sec:NonInt.}. Notably, we find that the interacting occupations
derived from Eqs.~\eqref{Eq:G<fte} and~\eqref{Eq:G<(1)te} can be also connected with the noninteracting
densities of Eqs.~\eqref{Eq:nf0} and~\eqref{Eq:n10}:
\begin{align}
&\left\langle n(t)\right\rangle^{f}=\frac{ 2\,\left\langle n(t)\right\rangle^{f}_{0}}{2+\left\langle n(t)\right\rangle^{f}_{0}-\left\langle n(t)\right\rangle^{f}_{0U}}\,,\label{Eq:nf}\\
&\left\langle n(t)\right\rangle^{(1)}=2\,\frac{\left\langle n(t)\right\rangle^{(1)}_{0}(2-\left\langle n(t)\right\rangle^{f}_{0U})+\left\langle n(t)\right\rangle^{(1)}_{0U} \left\langle n(t)\right\rangle^{f}_{0}}{(2+\left\langle n(t)\right\rangle^{f}_{0}-\left\langle n(t)\right\rangle^{f}_{0U})^2}\,,\label{Eq:n1}
\end{align}
where the subscript $U$ again designates the substitution $\veps_0\rightarrow \varepsilon_0+U$.
From the latter equations we can immediately derive the capacitive and dissipative currents,
\begin{align}
&I^{(1)}(t)\!=\!2\,\frac{I^{(1)}_{0}(t)(2-\left\langle n(t)\right\rangle^{f}_{0U}) +I^{(1)}_{0U}(t)\left\langle n(t)\right\rangle^{f}_{0}}{(2+\left\langle n(t)\right\rangle^{f}_{0}-\left\langle n(t)\right\rangle^{f}_{0U}) ^2}\,,\label{Eq:I1}\\
&I^{(2)}(t)\!=\! e\partial_t\left\langle n(t)\right\rangle^{(1)}\,,\label{Eq:I2}
\end{align}
These are the central results of our paper. In particular, Eq.~\eqref{Eq:I1} states that the
leading-order current for interacting electrons is given by a weighted sum of
the noninteracting expressions [Eq.~\eqref{Eq:I10}] corresponding
to two resonances, namely, $\veps_0$ and $\varepsilon_0+U$.
This finding is particularly appealing since it anticipates the main transformation
of the noninteracting results---the current pulses, for moderate values of $U$,
will split into two separate peaks. We will now confirm our expectation with
exact numerical results.

\begin{figure}[t]
\centering
\includegraphics[width=0.40\textwidth]{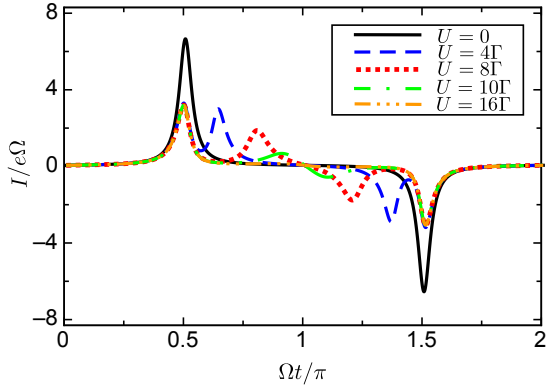}
\caption{Interacting charge current (Coulomb blockade regime) as a function of time for different
values of the charging energy $U$.
Parameters: $\veps_0=0$, $\varepsilon_{ac}=10\Gamma$, $\hbar\Omega=0.02\,\Gamma$, and $k_BT=0$.}
\label{Fig:I1CBUEac10}
\end{figure}

Figure~\ref{Fig:I1CBUEac10} shows the behavior of total charge current, $I^{(1)}(t)+I^{(2)}(t)$,
as a function of time for $\veps_0=0$, $\varepsilon_{ac}=10\Gamma$, and different values
of the charging energy $U$ at zero temperature. For $U=0$ (solid black line) we reproduce the curve from
Fig.~\ref{Fig:I1kBTEac10EacEd0} for comparison with the nonzero $U$ results.
Strikingly enough, for $U=4\Gamma$ (dashed blue line) both the peak and the dip split into two resonances each.
Therefore, we have {\em two} consecutive electron emissions (absorptions) whenever $\veps_0$
and $\veps_0 +U$ cross above (below) the lead Fermi level thus satisfying the resonant condition.
Furthermore, the amplitude of each resonance becomes reduced as compared with the noninteracting case.
This can be understood if one recalls that in the noninteracting case the dot level is spin-degenerate while
for interacting electrons each resonance can be occupied with at most one electron due to Pauli blocking.
The splitting gradually increases as $U$ is enhanced
[see the transition to the dotted red line ($U=8\Gamma$) and the dashed-dotted green line ($U=10\Gamma$)]
 because the second resonance shifts to higher (lower)
times as compared with the peak (dip) originally present for $U=0$. This second resonance
decreases its amplitude until it vanishes for $U>\veps_{ac}=10\Gamma$ (dashed-dotted orange curve).
This effect can be explained if we notice that the resonance $\veps_0+U$ never crosses
the Fermi level if $U>\veps_{ac}$. In other words, the two resonances can be occupied
(at least partially) only if $U<E_F+\veps_{ac}-\varepsilon_0$.

\begin{figure}[t]
\centering
\includegraphics[width=0.46\textwidth]{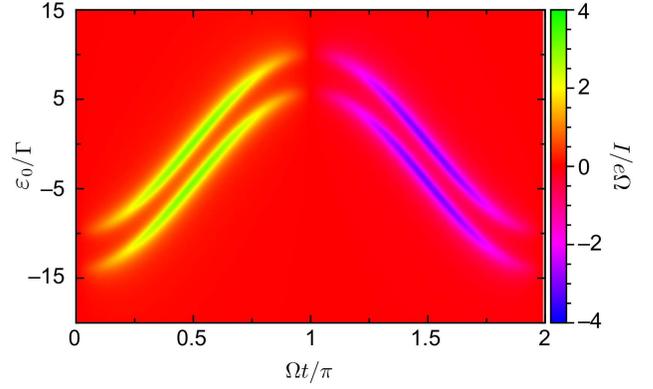}
\caption{Interacting charge current (Coulomb blockade regime) as a function of the dot energy level (vertical axis) and time (horizonal axis). Parameters: $\veps_{ac}=10\Gamma$, $U=4\Gamma$, $\hbar\Omega=0.02\,\Gamma$, and $k_BT=0$.}
\label{Fig:I1CBOtEdEac10}
\end{figure}

In Fig.~\ref{Fig:I1CBOtEdEac10} we present the total current as a function of time and the dot energy level position for a fixed
charging energy ($U=4\Gamma$) and ac amplitude ($\veps_{ac}=10\Gamma$).
We see clear signatures of the peak splitting for a wide range of energy levels since as we tune $\veps_0$ the resonant condition is satisfied at different times, as explained above. 

\begin{figure}[t]
\centering
\includegraphics[width=0.4\textwidth]{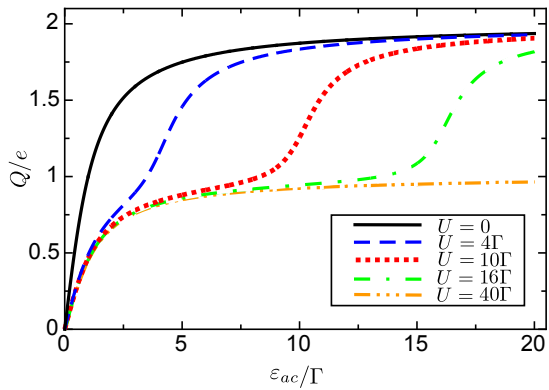}
\caption{Charge (Coulomb blockade regime) as a function of the ac amplitude $\veps_{ac}$. Parameters: $\varepsilon_{0}=0$, $\hbar\Omega=0.02\,\Gamma$, and $k_BT=0$.}
\label{Fig:QCBUEac}
\end{figure}

Importantly, electron-electron interactions affect the charge quantization in a mesoscopic capacitor.
From the total charge current we can obtain the charge $Q$ emitted for a half of a period
in terms of the occupation:
\begin{align}
Q=\int_0^{\tau/2} \!\! dt~ I(t) = e\,\big(\!\left\langle n(t=\tau/2)\right\rangle-\left\langle n(t=0)\right\rangle\big)\,,
\end{align}
where $\tau=2\pi/\Omega$ is the ac period and $\left\langle n(t)\right\rangle=\left\langle n(t)\right\rangle^f+\left\langle n(t)\right\rangle^{(1)}$ is the total occupation given by the sum of Eqs.~\eqref{Eq:nf} and~\eqref{Eq:n1}
to lowest order in frequency.
Figure~\ref{Fig:QCBUEac} shows $Q$ as a function of the ac amplitude for different values of the Coulomb strength, $U$. For $U=0$ we recover a full charge quantization at large values of the harmonic potential~\cite{mos08}. With increasing electron-electron interactions,
a new plateau emerges for intermediate values of $\veps_{ac}$. This phenomenon is exclusively due to Coulomb repulsion effects since when $U>\Gamma$ the dot energy level is split into two resonances, $\varepsilon_0$ and $\varepsilon_0+U$,
which are occupied sequentially as $\veps_{ac}$ grows. It is worth noting that the transition between plateaus
shifts to larger values of energy as $U$ increases because when $U>\veps_{ac}$ only the resonance at $\varepsilon_0$ is able to fulfill the resonant condition and the second plateau ceases to be visible. Therefore, it is crucial
to take into account electron-electron interactions to give precise predictions on the charge quantization amplitude
and its domain.

Let us turn now to the differential capacitance and resistance. In Eqs.~\eqref{Eq:C0t} and~\eqref{Eq:R0t} we obtained
their full expressions for noninteracting electrons. When interactions are present, we should combine
Eq.~\eqref{Eq:eI(t)} together with Eqs.~\eqref{Eq:I1} and~\eqref{Eq:I2} to arrive at the following relation:
\begin{align}
C_{\partial}(t)=2\,\frac{C^{0}_{\partial}(t)(2-\left\langle n(t)\right\rangle^{f}_{0U})+C^{0U}_{\partial}\!(t) \left\langle n(t)\right\rangle^{f}_{0}}{ (2+ \left\langle n(t)\right\rangle^{f}_{0}-\left\langle n(t)\right\rangle^{f}_{0U}) ^2}\,.\label{Eq:Ct}
\end{align}
Remarkably, we again find the nice result that the Coulomb-blockaded capacitance $C_{\partial}(t)$
can be written in terms of a weighted sum of noninteracting capacitances renormalized by interactions.
The weight factors depend themselves on shifted occupations calculated in the absence
($\left\langle n(t)\right\rangle^{f}_{0}$) and in the presence ($\left\langle n(t)\right\rangle^{f}_{0U}$)
of interactions. Nevertheless, the analytic expression for the resistance is too lengthy to be included here.
For the numerical calculations we shall use the definition
\begin{align}
R_{\partial}(t)=e\frac{I^{(2)}(t)}{C_{\partial}(t) \partial_t (C_{\partial}(t) (\partial_t \veps_{ac}(t))) }\,.\label{Eq:Rt}
\end{align}

\begin{figure}[t]
\centering
\includegraphics[width=0.40\textwidth]{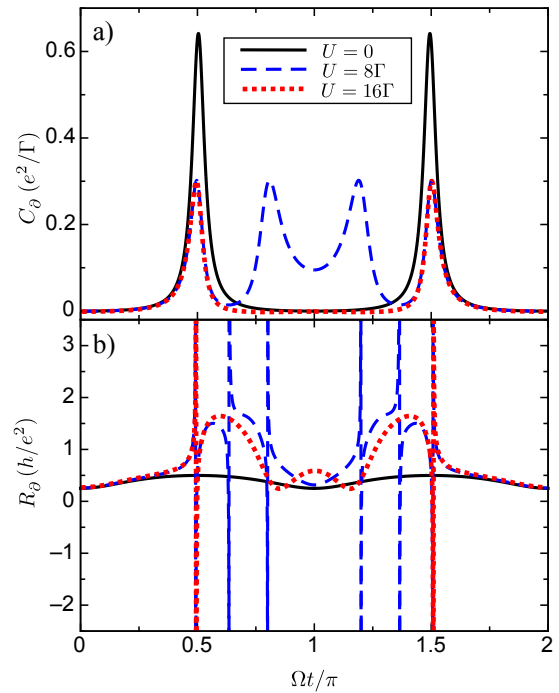}
\caption{Differential capacitance (a) differential resistance (b) as a function of time for different values of the charging energy $U$.
Parameters: $\veps_0=0$, $\varepsilon_{ac}=10\Gamma$, and $k_BT=0$.}
\label{Fig:CqRqCBUEac10}
\end{figure}

In Fig.~\ref{Fig:CqRqCBUEac10} we plot Eqs.~\eqref{Eq:Ct} and~\eqref{Eq:Rt} as a function of time for different Coulomb strengths.
In the top panel [Fig.~\ref{Fig:CqRqCBUEac10}(a)], we depict $C_{\partial}(t)$
in units of $e^2/\Gamma$. As expected, the capacitance, which mimics the instantaneous density of states,
undergoes a double splitting for finite charging energies (cf. the case $U=0$ showed in solid black line
with the case $U=8\Gamma$ in dashed blue line). The four-peak structure arises from multiple passings
(upward and downward) of the resonances $\veps_0$ and $\varepsilon_0+U$ across the Fermi energy.
Our calculations predict that four peaks (two in each half cycle) will appear in the Coulomb blockade regime
($U>\pi \Gamma$) and for sufficiently low temperature. Further increase of $U$ leads to a recovery of the two peaks but
with reduced amplitude. In general, for energies $U>E_F+\veps_{ac}-\varepsilon_0$
(with $\veps_0>0$) the resonance lying at $\varepsilon_0+U$ is not able to fulfill the resonant condition
and we recover the $U=0$ case but with half-height peaks due to the $1/2$ occupation (on average) of each spin level.

We show the differential resistance $R_{\partial}$ in Fig.~\ref{Fig:CqRqCBUEac10}(b).
Already for $U=0$ we find departures from the universal charge relaxation resistance value ${h}/{4 e^2}$.
These deviations are stronger as $U$ increases and lead to {\em negative} values of $R_{\partial}$
for certain values of time. Therefore, we cannot identify the product $C_\partial R_{\partial}$
with a delay time since this interpretation is physically meaningful in linear response only.
In fact, at some points the resistance diverges. Analogous resistance divergences have been
found in the thermoelectric transport~\cite{lim13} but here the effect is purely electric.
Equation~\eqref{Eq:Rt} dictates that the differential resistance is inversely proportional to the derivative of the differential
capacitance. As a consequence, $R_{\partial}$ diverges whenever this derivative vanishes. This implies
that the resistance divergences are correlated with the maxima or minima of $C_\partial$, as can be easily
inferred from a close inspection of Figs.~\ref{Fig:CqRqCBUEac10}(a) and~\ref{Fig:CqRqCBUEac10}(b).

\begin{figure}[t]
\centering
\includegraphics[width=0.40\textwidth]{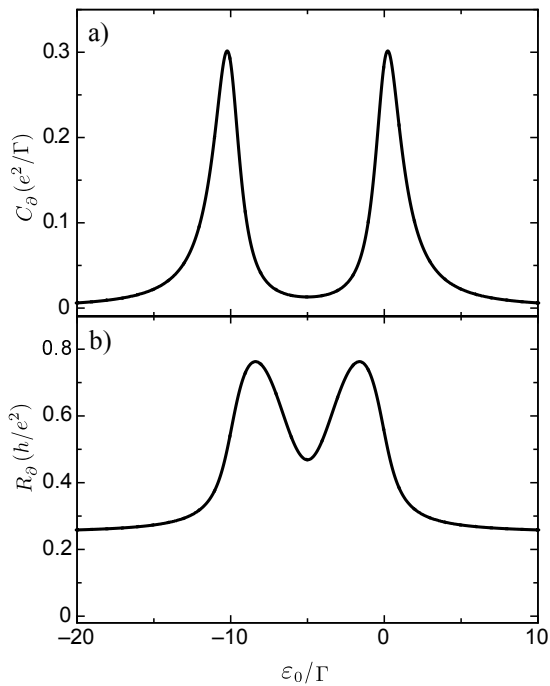}
\caption{Quantum capacitance (a) and charge relaxation resistance resistance (b) as a function of dot energy level in the linear regime, $\veps_{ac}\rightarrow 0$, and for interacting electrons in the Coulomb-blockade regime.
Parameters: $U=10\Gamma$, and $k_BT=0$.}
\label{Fig:CqRqCBVac0}
\end{figure}

A natural question is then whether the strong fluctuations of the nonlinear resistance away
from its quantized value persist in the linear regime. To examine this, we take the limit $\veps_{ac}\to 0$
in Eqs.~\eqref{Eq:Ct} and~\eqref{Eq:Rt}. We find for $k_BT =0$ the expressions
\begin{align}
C_{\partial}&= 4\,\frac{\mathcal{D} (2-\left\langle n\right\rangle^{f}_{0U}) +\mathcal{D}_U  \left\langle n\right\rangle^{f}_{0}}{ (2+\left\langle n\right\rangle^{f}_{0}-\left\langle n\right\rangle^{f}_{0U}) ^2}\,,\label{Eq:CtVac0}\\
R_{\partial}&=\frac{h }{8 e^2}\frac{\mathcal{D}^2 (2-\left\langle n\right\rangle^{f}_{0U}) +\mathcal{D}_U^2  \left\langle n\right\rangle^{f}_{0} }{(\mathcal{D} (2-\left\langle n\right\rangle^{f}_{0U}) +\mathcal{D}_U  \left\langle n\right\rangle^{f}_{0}  ) ^2}\nonumber\\
&\quad\quad\quad\quad\quad\quad\quad\quad\quad\times (2+ \left\langle n\right\rangle^{f}_{0}-\left\langle n\right\rangle^{f}_{0U})^2\,,\label{Eq:RtVac0}
\end{align}
where $\mathcal{D}=\Gamma/[(E_F-\veps_0)^2+\Gamma^2]$ and $\mathcal{D}_U=\Gamma/[(E_F-\varepsilon_0-U)^2+\Gamma^2]$. Interestingly, Eqs.~\eqref{Eq:CtVac0} and~\eqref{Eq:RtVac0}
depend on the mean frozen occupation. The capacitance is a weighted sum of densities of states
and will therefore show two peaks at $\veps_0\simeq E_F$ and $\varepsilon_0\simeq E_F-U$
[see Fig.~\ref{Fig:CqRqCBVac0}(a) where we depict the capacitance as a function of the dot level].
Even in the presence of interactions the capacitance can be traced back to a spectroscopic
measure of the dot spectral function. However, the charge relaxation
resistance is no longer constant as in the noninteracting case.
In Fig.~\ref{Fig:CqRqCBVac0}(b) we observe a strong energy dependence of $R_{\partial}$
with $\veps_0$. Only when the dot level is clearly off resonance (either $\varepsilon_0\gg\Gamma$
or $\veps_0\ll\Gamma$) do we recover the universal value $h/4e^2$.
In both cases the reason is clear---either for $\veps_0$ well above $E_F$ or for a deep level configuration,
interactions play no role and the noninterating result is restored. 
In the electron-hole symmetry point [$\veps_0=(E_F-U)/2$] the system behaves effectively as a single channel
conductor because the occupation per spin is $1/2$. For dot energies in between the electron-hole symmetry point
and the off-resonant situation, the charge relaxation resistance acquires its maximum value, which is 
sample dependent. We attribute this resistance increase to the maximal charge fluctuations that operate
around the point $\veps_0\simeq -\Gamma$ and its symmetric counterpart $\varepsilon_0\simeq U-\Gamma$.
We notice that significant enhancements of $R_\partial$ have been previously reported in the literature
for interacting $RC$ circuits~\cite{lee11,fil11}.

\section{Conclusions}\label{Sec:Conclusions}

In summary, we have investigated Coulomb blockade effects
in a coherent source of single-electrons driven by a monochromatic excitation.
Using a nonequilibrium Green's function approach valid for
arbitrarily large amplitudes of the ac potential,
we have found that the 
current peaks associated to electron emission and absorption
become split in the Coulomb blockade regime.
The effect is particularly intense for the emitted charge,
with additional quantization steps as a function
of the ac forcing. Our model is capable of describing
the noninteracting case ($U=0$) up to strong interactions ($U\to\infty$)
within the Coulomb blockade regime. While for $U=0$
our theory produces two-electron or two-hole pulses,
for $U\to\infty$ our model predicts single-electron or single-hole
pulses. For intermediate values of $U$ one may have
two single-electron or single-hole pulses separated in time.
Our model system is a mesoscopic capacitor but our results are equally
relevant for different single-electron sources such as those formed
with dopant atoms in silicon~\cite{lan12,roc13,tet14}
or dots embedded in coplanar cavities~\cite{del11,fre12,cot15,bru16}.

Further investigations should address the role of cotunneling
processes which are dominant in the Coulomb blockade valley
at temperatures $k_B T\ll \Gamma$. One possibility is to relax
the conditions given by Eqs.~\eqref{Eq:Aprox.1} and~\eqref{Eq:Aprox.2}
and to make a step further in the equation-of-motion hierarchy.
In particular, spin-flip cotunneling processes would lead to Kondo
correlations that would alter the picture discussed here.
In general, we expect the minimum between current peaks (dips)
to rise (lower) due to the buildup of a many-body Kondo resonance
pinned at the Fermi energy.
An additional peak should then appear in the quantum capacitance
since it is proportional to the local density of states. However, a new
energy scale ($k_B T_K$ with $T_K$ the Kondo temperature)
would arise and a more careful analysis should be carried out.

Another assumption of our model is the spin degeneracy
in both the dot level and the coupled reservoir [cf. Eq.~\eqref{Eq:nonmagnetic}].
Introducing a Zeeman splitting $\Delta_Z$ would lead to extra splittings
that would compete with the existing ones depending on the strength
of $\Delta_Z$ as compared with $\Gamma$, $U$ and $k_BT$.
We note that the original experiments by F\`eve \textit{et al.}~\cite{fev07}
applied a strong magnetic field that drove the system into the quantum Hall
regime. Moreover, the dot coupled to a gate with a large capacitance and
charging effects were then negligible. To test our predictions, we would need
a smaller dot in the absence of magnetic fields (or with Zeeman fields smaller
than the characteristic energy scales).

Finally, we have focused on the adiabatic regime (low frequencies).
This approximation is valid if one is interested in the capacitance and the charge
relaxation resistance. Arbitrary frequencies are beyond the scope of the present work
but are certainly interesting (for $U=0$ see, e.g., Refs.~\cite{bat11,bat12}). In fact, for larger frequencies
(larger than the GHz scale considered in this work)
photon-assisted tunneling takes place~\cite{kou94,kog04}
and our frequency expansion breaks down.
It would be highly desirable to take into account large frequencies and amplitudes
in a unified framework for the purely ac transport of electrons in nanostructures.

\acknowledgments
We thank A. Cottet, M. Moskalets and P. Samuelsson for useful comments.
This work has been supported by 
MINECO under Grant No.\ FIS2014-52564.

\appendix

\section{Fourier transform and mixed time-energy representation}\label{App:Fourier}

The double Fourier transformation and its inverse are defined as
\begin{align}
&\mathcal{G} (t,t')=\sum_{m,n} \int \frac{d\veps}{2 \pi}~e^{-i (\veps+m\hbar \Omega)t/\hbar}e^{i (\veps+n\hbar \Omega)t'/\hbar}\mathcal{G}(m-n,\veps_n)\label{EqA:gttptogmn} \,,
\\
&\mathcal{G}(m-n,\veps_n)=\int_0^{\tau}\!\!\frac{dt}{\tau}\!\int\!\!\frac{dt'}{\hbar}~e^{i (\veps+m\hbar \Omega)t/\hbar}e^{-i (\veps+n\hbar \Omega)t'/\hbar}\mathcal{G} (t,t')\label{EqA:gmntogttp} \,,
\end{align}
where $m$ and $n$ are intergers, $\tau=2 \pi/\Omega$ is the ac period, and $\veps_n = \veps + n\hbar\Omega$.
Notice that only the states whose energies differ by interger times $\hbar\Omega$ can be coupled.
It is convenient to employ the mixed time-energy representation
\begin{equation}\label{EqA:gtetogene}
\mathcal{G}(t,\veps)=\sum_n e^{-i n \Omega t}\mathcal{G}(n,\veps) \,.
\end{equation}
The Fourier transform can then be written in the form
\begin{equation}\label{EqA:gttptogte}
\mathcal{G}(t,t')=\int\frac{d\veps}{2 \pi}~e^{-i\veps(t-t')/\hbar}\mathcal{G}(t,\veps)
\end{equation}
and the corresponding inverse Fourier transforms are given by
\begin{align}
&\mathcal{G}(t,\veps)=\int\frac{dt'}{\hbar}~e^{i \varepsilon(t-t')/\hbar}\mathcal{G}(t,t')\label{EqA:gtetottp} \,,
\\
&\mathcal{G}(n,\veps)=\int_0^{\tau}\frac{dt}{\tau}~e^{in\Omega t}\mathcal{G}(t,\veps)\label{EqA:genetogte} \,,
\end{align}
respectively.

\section{Frequency expansion}

We begin by applying the double Fourier transform Eq.~\eqref{EqA:gmntogttp}
to Eqs.~\eqref{Eq:G} and~\eqref{Eq:Corr}:
\begin{align}
&G_{\sm}(n,\veps)=\ttg_{\sm}(n,\veps)+\sum_p G_{\sm}(n-p,\veps_p)[\veps_{ac}\ttg_{\sm}](p,\veps)\nonumber
\\
&\quad\quad\quad\quad+\sum_{p,q}G_{\sm}(n-p,\veps_p) \Sigma_0(p-q,\veps_q) \ttg_{\sm}(q,\veps)\nonumber
\\
&\quad\quad\quad\quad\quad\quad+U\sum_p\langle\langle d_{\sm},d_{\sm}^{\dag}n_{\bar{\sm}}\rangle\rangle(n-p,\veps_p) \ttg_{\sm}(p,\veps) \,,\label{EqA:Gne}
\\
&\langle\langle d_{\sm},d_{\sm}^{\dag}n_{\bar{\sm}}\rangle\rangle(n,\veps)=\left[\left\langle n_{\bar{\sm}}\right\rangle\!\ttg_{\sm}\right](n,\veps)\nonumber
\\
&\quad\quad\quad\quad+\sum_p \langle\langle d_{\sm},d_{\sm}^{\dag}n_{\bar{\sm}}\rangle\rangle(n-p,\veps_p)\left[\veps_{ac}\ttg_{\sm}\right](p,\veps)\nonumber
\\
&\quad\quad+\sum_{p,q} \langle\langle d_{\sm},d_{\sm}^{\dag}n_{\bar{\sm}}\rangle\rangle(n-p,\veps_p) \Sigma_0(p-q,\veps_q) \ttg_{\sm}(q,\veps)\nonumber
\\
&\quad\quad\quad\quad\quad\quad+\,U\sum_p \langle\langle d_{\sm},d_{\sm}^{\dag}n_{\bar{\sm}}\rangle\rangle(n-p,\veps_p) \ttg_{\sm}(p,\veps)\,.\label{EqA:Corrne}
\end{align}

The retarded/advanced and lesser Green's functions then follow from Eqs.~\eqref{EqA:Gne} and \eqref{EqA:Corrne} by applying the Langreth's rules~\cite{jauho}.

\subsection{Noninteracting case}\label{App:Serie.0}

In the noninteracting case, we set $U=0$ and therefore $\langle\langle d_{\sm},d_{\sm}^{\dag}n_{\bar{\sm}}\rangle\rangle(n,\veps)$ in Eqs.~\eqref{EqA:Gne} and~\eqref{EqA:Corrne} is neglected.

\subsubsection{Retarded and advanced Green's function}

The retarded/advanced dot Green's function is given by
\begin{multline}\label{EqA:Gr0ne.1}
\mathcal{G}^{r/a}_{\sm}(n,\veps)=\ttg^{r/a}_{\sm}(n,\veps)+\sum_p \mathcal{G}^{r/a}_{\sm}(n-p,\veps_p)[\veps_{ac}\ttg^{r/a}_{\sm}](p,\veps)
\\
+\sum_{p,q}\mathcal{G}^{r/a}_{\sm}(n-p,\veps_p) \Sigma^{r/a}_0(p-q,\veps_q) \ttg^{r/a}_{\sm}(q,\veps)
\end{multline}
with
\begin{subequations}\label{EqA:Exp1}
\begin{align}
&\ttg^{r/a}_{\sm}(n,\veps)=\frac{\delta_{n,0}}{\veps-\veps_{\sm}\pm i0^+}=\delta_{n,0}\ttg^{r/a}_{\sm}(\veps) \,,
\\
&\Sigma^{r/a}_0(m-n,\veps_n)=\mp i\delta_{m,n}\Gamma(\veps_n)=\delta_{m,n}\Sigma^{r/a}_0(\veps_n) \,,
\\
&[\veps_{ac}\ttg^{r/a}_{\sm}](n,\veps)=\frac{\veps_{ac}}{2}(\delta_{n,1}+\delta_{n,-1})\ttg^{r/a}_{\sm}(\veps) \,,
\end{align}
\end{subequations}
where $\Gamma(\veps_n)=2 \pi |V_k|^2 \rho(\veps_n)$ and $\rho(\veps_n)=\sum_k\delta(\veps_n-\veps_k)$ is the reservoir density of states.

Introducing Eq.~\eqref{EqA:Exp1} into Eq.~\eqref{EqA:Gr0ne.1}, we find
\begin{multline}\label{EqA:Gr0ne.2}
\mathcal{G}^{r/a}_{\sm}(n,\veps)=\Big(\delta_{n,0}+\frac{\veps_{ac}}{2}\sum_{p=\pm 1} \mathcal{G}^{r/a}_{\sm}(n-p,\veps_p)\Big)\mathscr{G}^{r/a}_{\sm}(\veps) \,,
\end{multline}
where
\begin{align}\label{EqA:mathscrG}
\mathscr{G}^{r/a}_{\sm}(\veps)=\frac{1}{\veps-\veps_{\sm}-\Sigma^{r/a}_0(\veps)} \,.
\end{align}

We now expand in powers of $\hbar\Omega$
\begin{multline}\label{EqA:Frec}
\mathcal{G}(n-p,\veps_p)=\mathcal{G}^f(n-p,\varepsilon)\\
+\hbar\Omega\Big(p\partial_{\veps}\mathcal{G}^f(n-p,\varepsilon)+\mathcal{G}^{(1)}(n-p,\varepsilon)\Big) +\dots\,,
\end{multline}
and substitute it in Eq.~\eqref{EqA:Gr0ne.2} to find
\begin{align}
&\mathcal{G}^{r/a,f}_{\sm}(n,\veps)= \Big(\delta_{n,0}+\frac{\veps_{ac}}{2}\!\sum_{p=\pm 1}\!\mathcal{G}^{r/a}_{\sm}(n-p,\veps)\Big)\mathscr{G}^{r/a}_{\sm}(\veps) \,,\label{EqA:Grf0ne} 
\\
&\mathcal{G}^{r/a,(1)}_{\sm}(n,\veps)=\frac{\veps_{ac}}{2}\!\sum_{p=\pm 1}\!\Big( p\partial_{\veps}\mathcal{G}^{r/a,f}_{\sm}(n-p,\veps) \nonumber\\
&\quad\quad\quad\quad\quad\quad\quad\quad\quad\quad+\mathcal{G}^{r/a,(1)}_{\sm}(n-p,\veps) \Big)\mathscr{G}^{r/a}_{\sm}(\veps) \,.\label{EqA:Gr(1)0ne} 
\end{align}
Using Eq.~\eqref{EqA:gtetogene} and taking into account the
wide band limit $\Gamma(\veps)=\Gamma$, which is a good approximation for reservoirs with
flat densities of states, we arrive at Eqs.~\eqref{Eq:Grf0} and~\eqref{Eq:Gr(1)0} of the main text.

\subsubsection{Lesser Green's function}

The lesser Green's function for the quantum dot electrons can be obtained as
\begin{widetext}
\begin{multline}\label{EqA:G<0ne.1}
\mathcal{G}^<_{\sm}(n,\veps)=\delta_{n,0}\ttg^<_{\sigma}(n,\varepsilon)
+\sum_p \Big(\mathcal{G}^r_{\sm}(n-p,\veps_p)[\varepsilon_{ac}\ttg^<_{\sigma}](p,\varepsilon) 
+\mathcal{G}^<_{\sm}(n-p,\veps_p)[\varepsilon_{ac}\ttg^a_{\sigma}](p,\varepsilon)\Big)
+\sum_{p,q}\Big( \mathcal{G}^r_{\sm}(n-p,\veps_p) \Sigma^r_0(p-q,\varepsilon_q) \ttg^<_{\sigma}(q,\varepsilon)\\
+\mathcal{G}^r_{\sm}(n-p,\veps_p) \Sigma^<_0(p-q,\varepsilon_q) \ttg^a_{\sigma}(q,\varepsilon)+\mathcal{G}^<_{\sigma}(n-p,\varepsilon_p) \Sigma^a_0(p-q,\varepsilon_q) \ttg^a_{\sigma}(q,\varepsilon)\Big) \,,
\end{multline}
\end{widetext}
where
\begin{subequations}\label{EqA:Exp2}
\begin{align}
&\ttg^<_{\sm}(n,\veps)=2 \pi i\delta_{n,0} \delta(\veps-\veps_{\sm})f(\veps_{\sm})=\delta_{n,0}\ttg^<_{\sm}(\veps) \,,
\\
&\Sigma^<_{0}(m-n,\veps_n)=2i \delta_{m,n} \Gamma(\veps_n) f(\veps_n)= \delta_{m,n}\Sigma^<_{0}(\veps_n) \,,
\\
&[\veps_{ac} \ttg^<_{\sm}](n,\veps)=\frac{\veps_{ac}}{2}(\delta_{n,1}+\delta_{n,-1})\ttg^<_{\sm}(\veps) \,.
\end{align}
\end{subequations}
Introducing Eqs.~\eqref{EqA:Exp1} and~\eqref{EqA:Exp2} into Eq.~\eqref{EqA:G<0ne.1} and using $\ttg^{r,-1}_{\sm}(\veps)\ttg^<_{\sm}(\veps)=0$ we find
\begin{multline}\label{EqA:G<0ne.2}
\mathcal{G}^{<}_{\sm}(n,\veps)=\Big(\frac{\veps_{ac}}{2}\sum_{p=\pm 1}\mathcal{G}^{<}_{\sm}(n-p,\veps_p)
\\
+\mathcal{G}^{r}_{\sm}(n,\veps) \Sigma^<_0(\varepsilon)\Big)\mathscr{G}^a_{\sigma}(\veps) \,.
\end{multline}
This is the starting point for a series expansion in powers of $\hbar\Omega$. The procedure is analogous to
Eq.~\eqref{EqA:Frec}. Then, the frozen and first order terms in $\Omega$ become, respectively,
\small{\begin{align}
&\mathcal{G}^{<,f}_{\sm}(n,\veps)=\Big(\frac{\veps_{ac}}{2}\sum_{p=\pm 1}\mathcal{G}^{<,f}_{\sm}(n-p,\veps)\nonumber
\\
&\quad\quad\quad\quad\quad\quad\quad\quad\quad\quad\quad+\mathcal{G}^{r,f}_{\sm}(n,\veps) \Sigma^<_0(\veps)\Big)\mathscr{G}^a_{\sm}(\veps) \,,\label{EqA:G<f0ne}
\\
&\mathcal{G}^{<,(1)}_{\sm}(n,\veps)=\bigg(\frac{\veps_{ac}}{2}\sum_{p=\pm 1}\Big(p\partial_{\veps}\mathcal{G}^{<,f}_{\sm}(n-p,\veps)\nonumber
\\
&\quad\quad+\mathcal{G}^{r,(1)}_{\sm}(n-p,\veps)\Big)+\mathcal{G}^{r,(1)}_{\sm}(n,\veps)\Sigma^<_0(\veps)\bigg)\mathscr{G}^a_{\sm}(\veps) \,.\label{EqA:G<(1)0ne}
\end{align}}
As discussed earlier, we can again use Eq.~\eqref{EqA:gtetogene} and consider the wide band limit,
which leads to Eqs.~\eqref{Eq:G<f0} and~\eqref{Eq:G<(1)0}.

\subsection{Interacting case (Coulomb blockade regime)}\label{App:Serie.CB}

In order to describe the Coulomb blockade regime, we consider the nonzero $U$ case.
Hence, Eq.~\eqref{EqA:Corrne} must be taken into account.

\subsubsection{Retarded and advanced Green's function}

The retarded/advanced Green's functions are simply derived from Eq.~\eqref{EqA:Gne} and~\eqref{EqA:Corrne},
yielding
\begin{align}
&{{G}}^{r/a}_{\sm}(n,\veps)={\ttg}^{r/a}_{\sm}(n,\veps)+\sum_p {{G}}^{r/a}_{\sm}(n-p,\veps_p)[\veps_{ac}{\ttg}^{r/a}_{\sm}](p,\veps)\nonumber
\\
&\quad\quad\quad+\sum_{p,q}{{G}}^{r/a}_{\sm}(n-p,\veps_p) {\Sigma}^{r/a}_0(p-q,\veps_q) {\ttg}^{r/a}_{\sm}(q,\veps)\nonumber
\\
&\quad\quad\quad\quad+U\sum_p\langle\langle d_{\sm},d_{\sm}^{\dag}n_{\bar{\sm}}\rangle\rangle^{r/a}(n-p,\veps_p) {\ttg}^{r/a}_{\sm}(p,\veps) \,,\label{EqA:Grne.1}
\end{align}
\begin{align}
&\langle\langle d_{\sm},d_{\sm}^{\dag}n_{\bar{\sm}}\rangle\rangle^{r/a}(n,\veps)=\left[\left\langle n_{\bar{\sm}}\right\rangle\!{\ttg}^{r/a}_{\sm}\right](n,\veps)\nonumber
\\
&\quad\quad\quad+\sum_p \langle\langle d_{\sm},d_{\sm}^{\dag}n_{\bar{\sm}}\rangle\rangle^{r/a}(n-p,\veps_p)\left[\veps_{ac}{\ttg}^{r/a}_{\sm}\right](p,\veps)\nonumber
\\
&\quad+\sum_{p,q} \langle\langle d_{\sm},d_{\sm}^{\dag}n_{\bar{\sm}}\rangle\rangle^{r/a}(n-p,\veps_p) {\Sigma}^{r/a}_0(p-q,\veps_q) {\ttg}^{r/a}_{\sm}(q,\veps)\nonumber
\\
&\quad\quad\quad\quad+\,U\sum_p \langle\langle d_{\sm},d_{\sm}^{\dag}n_{\bar{\sm}}\rangle\rangle^{r/a}(n-p,\veps_p) {\ttg}^{r/a}_{\sm}(p,\veps)\,,\label{EqA:Corrrne.1}
\end{align}
with
\begin{align}\label{EqA:Exp3}
\left[\left\langle n_{\bar{\sm}}\right\rangle\!{\ttg}^{r/a}_{\sm}\right](n,\veps)=\left\langle n_{\bar{\sm}}\right\rangle_n{\ttg}^{r/a}_{\sm}(\veps)\,.
\end{align}
where we have used the Fourier expansion
\begin{equation}
\left\langle n_{\bar{\sm}}(t)\right\rangle=\sum_n\left\langle n_{\bar{\sm}}\right\rangle_n e^{-in\Omega t} \,.
\end{equation}
We substitute Eqs.~\eqref{EqA:Exp1} and~\eqref{EqA:Exp3} into Eqs.~\eqref{EqA:Grne.1} and~\eqref{EqA:Corrrne.1} and find
\begin{align}
&{G}^{r/a}_{\sm}(n,\veps)=\Big( \delta_{n,0}+\frac{\veps_{ac}}{2}\sum_{p=\pm 1} {G}^{r/a}_{\sm}(n-p,\veps_p)  \nonumber
\\
&\quad\quad\quad\quad\quad+U\langle\langle d_{\sm},d_{\sm}^{\dag}n_{\bar{\sm}}\rangle\rangle^{r/a}(n,\veps) {\ttg}^{r/a}_{\sm}(\veps)\Big) \mathscr{G}^{r/a}_{\sm}(\veps)\,,\label{EqA:Grne.2}
\\
&\langle\langle d_{\sm},d_{\sm}^{\dag}n_{\bar{\sm}}\rangle\rangle^{r/a}(n,\veps) =\Big(\left\langle n_{\bar{\sm}}\right\rangle_n\nonumber
\\
&\quad+\frac{\veps_{ac}}{2}\sum_{p=\pm 1}\langle\langle d_{\sm},d_{\sm}^{\dag}n_{\bar{\sm}}\rangle\rangle^{r/a}(n-p,\veps_p)\Big) \mathscr{G}^{r/a}_{\sm}(\veps-U)\,,\label{EqA:Corrrne.2}
\end{align}
where $\mathscr{G}^{r/a}_{\sigma}$ is given by~\eqref{EqA:mathscrG}. The solution has poles at $\veps_{\sm}$ and $\veps_{\sm} + U$ such that it properly describes the Coulomb blockade.

Let us now expand in powers of $\hbar\Omega$. The expansion is based upon Eq.~\eqref{EqA:Frec}, which leads to
\begin{align}
&{G}^{r/a,f}_{\sm}(n,\veps)=\Big(\delta_{n,0}+\frac{\veps_{ac}}{2}\!\sum_{p=\pm 1}\!{G}^{r/a}_{\sm}(n-p,\veps)\nonumber
\\
&\quad\quad\quad\quad\quad\quad+U\langle\langle d_{\sm},d_{\sm}^{\dag}n_{\bar{\sm}}\rangle\rangle^{r/a,f}(n,\veps)\Big)\mathscr{G}^{r/a}_{\sm}(\veps)\,,\label{EqA:Grfne} 
\\
&{G}^{r/a,(1)}_{\sm}\,(n,\veps)=\Big(\frac{\veps_{ac}}{2}\! \sum_{p=\pm 1}\!\Big[ p\partial_{\veps}{G}^{r/a,f}_{\sm}(n-p,\veps)  \nonumber
\\
&\quad\quad\quad\quad\quad\quad\quad\quad\quad\quad\quad\quad+{G}^{r/a,(1)}_{\sm}(n-p,\veps) \Big]\nonumber
\\
&\quad\quad\quad\quad\quad\quad+U\langle\langle d_{\sm},d_{\sm}^{\dag}n_{\bar{\sm}}\rangle\rangle^{r/a,(1)}(n,\veps)\Big)\mathscr{G}^{r/a}_{\sm}(\veps) \,,\label{EqA:Gr(1)ne} 
\end{align}
where
\begin{align}
&\langle\langle d_{\sm},d_{\sm}^{\dag}n_{\bar{\sm}}\rangle\rangle^{r/a,f}(n,\veps)=\Big(\left\langle n_{\bar{\sm}}\right\rangle^f_n \nonumber
\\
&\quad+\frac{\veps_{ac}}{2}\!\sum_{p=\pm 1}\!\langle\langle   d_{\sm},d_{\sm}^{\dag}n_{\bar{\sm}}\rangle\rangle^{r/a,f}(n-p,\veps)\Big)\mathscr{G}^{r/a}_{\sm}(\veps-U)\,,\label{EqA:Corrrfne} 
\\
&\langle\langle  d_{\sm},d_{\sm}^{\dag}n_{\bar{\sm}}\rangle\rangle^{r/a,(1)}(n,\veps)=\Big(\left\langle n_{\bar{\sm}}\right\rangle^{(1)}_n \nonumber
\\
&\quad\quad\quad\quad+\frac{\veps_{ac}}{2}\! \sum_{p=\pm 1}\!\Big[ p\partial_{\veps}\langle\langle d_{\sm},d_{\sm}^{\dag}n_{\bar{\sm}}\rangle\rangle^{r/a,f}(n-p,\veps)\nonumber
\\
&\quad\quad\quad+\langle\langle  d_{\sm},d_{\sm}^{\dag}n_{\bar{\sm}}\rangle\rangle^{r/a,(1)}(n-p,\veps) \Big]\Big)\mathscr{G}^{r/a}_{\sm}(\veps-U) \,.\label{EqA:Corrr(1)ne} 
\end{align}
Expressing Eqs.~\eqref{EqA:Grfne} and~\eqref{EqA:Gr(1)ne} in the mixed time-energy representation leads to Eqs.~\eqref{Eq:Grfte} and Eqs.~\eqref{Eq:Gr(1)te}.

\subsubsection{Lesser Green's function}

Applying Langreth's rules again, the lesser Green's functions become
\begin{widetext}
\begin{multline}\label{EqA:G<ne.1}
{G}^<_{\sm}(n,\veps)=\delta_{n,0}\ttg^<_{\sm}(n,\veps)
+\sum_p \Big({G}^r_{\sm}(n-p,\veps_p)[\veps_{ac}\ttg^<_{\sm}](p,\veps) +{G}^<_{\sm}(n-p,\veps_p)[\veps_{ac}\ttg^a_{\sm}](p,\veps)\Big)
+U\sum_p\Big(\langle\langle d_{\sm},d_{\sm}^{\dag}n_{\bar{\sm}}\rangle\rangle^{r}(n-p,\veps_p)\ttg^<_{\sm}(p,\veps)
\\
+\langle\langle d_{\sm},d_{\sm}^{\dag}n_{\bar{\sm}}\rangle\rangle^{<}(n-p,\veps_p)\ttg^a_{\sm}(p,\veps)\Big)
+\sum_{p,q}\Big( {G}^r_{\sm}(n-p,\veps_p) \Sigma^r_0(p-q,\veps_q) \ttg^<_{\sm}(q,\veps)+{G}^r_{\sm}(n-p,\veps_p) \Sigma^<_0(p-q,\veps_q) \ttg^a_{\sm}(q,\veps)
\\+{G}^<_{\sm}(n-p,\veps_p) \Sigma^a_0(p-q,\veps_q) \ttg^a_{\sm}(q,\veps)\Big) \,,
\end{multline}
\begin{multline}\label{EqA:Corr<ne.1}
\langle\langle d_{\sm},d_{\sm}^{\dag}n_{\bar{\sm}}\rangle\rangle^<_{\sm}(n,\veps)
=\left[\left\langle n_{\bar{\sm}}\right\rangle\!{\ttg}^{<}_{\sm}\right](n,\veps)+\sum_p \Big(\langle\langle d_{\sm},d_{\sm}^{\dag}n_{\bar{\sm}}\rangle\rangle^r_{\sm}(n-p,\veps_p)[\veps_{ac}\ttg^<_{\sm}](p,\veps) +\langle\langle d_{\sm},d_{\sm}^{\dag}n_{\bar{\sm}}\rangle\rangle^<_{\sm}(n-p,\veps_p)[\veps_{ac}\ttg^a_{\sm}](p,\veps)\Big)
\\
+U\sum_p\Big(\langle\langle d_{\sm},d_{\sm}^{\dag}n_{\bar{\sm}}\rangle\rangle^{r}(n-p,\veps_p)\ttg^<_{\sm}(p,\veps)+\langle\langle d_{\sm},d_{\sm}^{\dag}n_{\bar{\sm}}\rangle\rangle^{<}(n-p,\veps_p)\ttg^a_{\sm}(p,\veps)\Big)
+\sum_{p,q}\Big( \langle\langle d_{\sm},d_{\sm}^{\dag}n_{\bar{\sm}}\rangle\rangle^r_{\sm}(n-p,\veps_p) \Sigma^r_0(p-q,\veps_q) \ttg^<_{\sm}(q,\veps)
\\+\langle\langle d_{\sm},d_{\sm}^{\dag}n_{\bar{\sm}}\rangle\rangle^r_{\sm}(n-p,\veps_p) \Sigma^<_0(p-q,\veps_q) \ttg^a_{\sm}(q,\veps)
+\langle\langle d_{\sm},d_{\sm}^{\dag}n_{\bar{\sm}}\rangle\rangle^<_{\sm}(n-p,\veps_p) \Sigma^a_0(p-q,\veps_q) \ttg^a_{\sm}(q,\veps)\Big) \,,
\end{multline}
\end{widetext}
with
\begin{align}\label{EqA:Exp4}
\left[\left\langle n_{\bar{\sm}}\right\rangle\!{\ttg}^{<}_{\sm}\right](n,\veps)=\left\langle n_{\bar{\sm}}\right\rangle_n{\ttg}^{<}_{\sm}(\veps) \,.
\end{align}
Inserting Eqs.~\eqref{EqA:Exp1}, \eqref{EqA:Exp2}, and \eqref{EqA:Exp4} into Eqs.~\eqref{EqA:G<ne.1} and~\eqref{EqA:Corr<ne.1}, and recalling that $\ttg^{r-1}_{\sm}(\veps)\ttg^<_{\sm}(\veps)=0$, we get
\begin{align}
&{G}^<_{\sm}(n,\veps)=\Big(\frac{\veps_{ac}}{2}\sum_{p=\pm 1} {G}^<_{\sm}(n-p,\veps_p)\nonumber\\
&\quad+U\langle\langle d_{\sm},d_{\sm}^{\dag}n_{\bar{\sm}}\rangle\rangle^{<}(n,\veps)+{G}^r_{\sm}(n,\veps) \Sigma^<_0(\veps) \Big) \mathscr{G}^a_{\sm}(\veps)\,,\label{EqA:G<ne.2}
\end{align}
\begin{align}
&\langle\langle d_{\sm},d_{\sm}^{\dag}n_{\bar{\sm}}\rangle\rangle^<_{\sm}(n,\veps)=\Big(\frac{\veps_{ac}}{2}\sum_{p=\pm 1} \langle\langle d_{\sm},d_{\sm}^{\dag}n_{\bar{\sm}}\rangle\rangle^<_{\sm}(n-p,\veps_p)\nonumber
\\
&\quad\quad\quad\quad\quad+\langle\langle d_{\sm},d_{\sm}^{\dag}n_{\bar{\sm}}\rangle\rangle^r_{\sm}(n,\veps) \Sigma^<_0(\veps) \Big) \mathscr{G}^a_{\sm}(\veps-U)\,.\label{EqA:Corr<ne.2}
\end{align}

The expansion in $\Omega$ yields
\begin{align}
&{G}^{<,f}_{\sm}(n,\veps)=\bigg(\frac{\veps_{ac}}{2}\sum_{p=\pm 1}{G}^{<,f}_{\sm}(n-p,\veps)\nonumber
\\
&+U\langle\langle d_{\sm},d_{\sm}^{\dag}n_{\bar{\sm}}\rangle\rangle^{<,f}_{\sm}(n,\veps)+{G}^{r,f}_{\sm}(n,\veps) \Sigma^<_0(\veps)\bigg)\mathscr{G}^a_{\sm}(\veps) \,,\label{EqA:G<fne}
\end{align}
\begin{align}
&{G}^{<,(1)}_{\sm}(n,\veps)=\bigg(\mathcal{G}^{r,(1)}_{\sm}(n,\veps)\Sigma^<_0(\veps)\nonumber
\\
&\quad\quad+\frac{\veps_{ac}}{2}\sum_{p=\pm 1}\Big[p\partial_{\veps}{G}^{<,f}_{\sm}(n-p,\veps)+\mathcal{G}^{r,(1)}_{\sm}(n-p,\veps)\Big]\nonumber
\\
&\quad\quad\quad\quad\quad\quad\quad+U\langle\langle d_{\sm},d_{\sm}^{\dag}n_{\bar{\sm}}\rangle\rangle^{<,(1)}_{\sm}(n,\veps)\bigg)\mathscr{G}^a_{\sm}(\veps) \,,\label{EqA:G<(1)ne}
\end{align}
and
\begin{align}
&\langle\langle d_{\sm},d_{\sm}^{\dag}n_{\bar{\sm}}\rangle\rangle^<_{\sm}(n,\veps)=\Big(\frac{\veps_{ac}}{2}\sum_{p=\pm 1} \langle\langle d_{\sm},d_{\sm}^{\dag}n_{\bar{\sm}}\rangle\rangle^{<,f}_{\sm}(n-p,\veps)\nonumber
\\
&\quad\quad\quad\quad+\langle\langle d_{\sm},d_{\sm}^{\dag}n_{\bar{\sm}}\rangle\rangle^{r,f}_{\sm}(n,\veps) \Sigma^<_0(\veps) \Big) \mathscr{G}^a_{\sm}(\veps-U)\,,\label{EqA:Corr<fne}
\end{align}
\begin{align}
&\langle\langle d_{\sm},d_{\sm}^{\dag}n_{\bar{\sm}}\rangle\rangle^{<,(1)}_{\sm}(n,\veps)=\Big(\langle\langle d_{\sm},d_{\sm}^{\dag}n_{\bar{\sm}}\rangle\rangle^{r,(1)}_{\sm}(n,\veps)\Sigma^<_0(\veps)\nonumber
\\
&\quad\quad\quad\quad\quad\quad+\frac{\veps_{ac}}{2}\sum_{p=\pm 1}\Big[p\partial_{\veps}\langle\langle d_{\sm},d_{\sm}^{\dag}n_{\bar{\sm}}\rangle\rangle^{<,f}_{\sm}(n-p,\veps)\nonumber
\\
&\quad\quad\quad+\langle\langle d_{\sm},d_{\sm}^{\dag}n_{\bar{\sm}}\rangle\rangle^{r,(1)}_{\sm}(n-p,\veps)\Big]\Big)\mathscr{G}^a_{\sm}(\veps-U) \,.\label{EqA:Corr<(1)ne}
\end{align}
Equations~\eqref{Eq:G<fte} and \eqref{Eq:G<(1)te} then follow easily.

\end{document}